\documentclass[twocolumn,secnumarabic,amssymb, nobibnotes, aps, superscriptaddress]{revtex4-2}
\usepackage[colorlinks=true, citecolor=blue, urlcolor=blue, linkcolor=blue, anchorcolor=blue]{hyperref}
\usepackage{graphicx}
\usepackage{bm}
\begin{document}

\title{Effect of strain on charge density wave order in $\alpha$-U}

\author{Liuhua Xie}
\affiliation{School of Physical Science and Technology, Southwest University, Chongqing 400715, China}
\affiliation{Institute of Materials, China Academy of Engineering Physics, Mianyang 621907, Sichuan, China}

\author{Hongkuan Yuan}
\email{yhk10@swu.edu.cn}
\affiliation{School of Physical Science and Technology, Southwest University, Chongqing 400715, China}

\author{Ruizhi Qiu}
\email{qiuruizhi@caep.cn}
\affiliation{Institute of Materials, China Academy of Engineering Physics, Mianyang 621907, Sichuan, China}

\date{\today}

\begin{abstract}
The effect of strain on charge density wave (CDW) order in $\alpha$-U is investigated within the framework of relativistic density-functional theory.
The energetical stability of $\alpha$-U with CDW distortion is enhanced by the tensile strain along $a$ and $b$ directions, which is similar to the case of negative pressure and normal.
However, the tensile strain along $c$ direction suppresses the energetical stability of CDW phase.
This abnormal effect could be understood from the emergence of a new one-dimensional atomic chain along $c$ direction in $\alpha$-U.
Furthermore, this effect is supported by the calculations of Fermi surface and phonon mode, in which the topological objects and the dynamical instability show opposite behavior between strain along $a$/$b$ and $c$ directions.
\end{abstract}
\maketitle

\section{Introduction}

Charge density wave (CDW), which is a simple periodic reorganization of electronic charge accompanied by a periodic modulation of the atomic structure, is still poorly understood and continue to bear surprises~\cite{Peierls1930,Froehlich1954,Gruener1988,Lander1994,Xi2015,Tonouchi2017,Manzeli2017,Gruner2017,Gao2018}.
This macroscopic quantum state occurs in a variety of materials including quasi-one-dimensional organic chain TTF-TCNG~\cite{Tonouchi2017}, layered transition-metal chalcogenide~\cite{Manzeli2017,Gruener1988}, and in particular, three-dimensional elemental metal uranium~\cite{Lander1994}.
Uranium is the \emph{only} element in the periodic table to exhibit a phase transition to CDW states at ambient pressure~\cite{Lander1994}.
The existence of these low-temperature CDW states in $\alpha$-uranium is well-established after several decades of thorough experimental work including the measurement of elastic constants~\cite{Fisher1961}, lattice parameters~\cite{Barrett1963,Steinitz1970}, phonon dispersion~\cite{Crummett1979,Smith1980}, and neutron back-reflection~\cite{Marmeggi1980}.
The transitions into CDW states take place at 43, 37, and 22 K, and then the corresponding CDW phases are denoted as $\alpha_1$, $\alpha_2$, and $\alpha_3$~\cite{Lander1994}.
The first transition from $\alpha$ into $\alpha_1$, involving only a doubling of the conventional unit cell of $\alpha$-U along the $a$ direction, is very simple and attracts considerable attention~\cite{Fast1998,Raymond2011,Dewaele2013,Qiu2016,Migdal2018}.

Since the effect of strain on the CDW materials could in principles induces transitions into a different order, the relevant studies increase significantly in the past decade~\cite{Springell2014,Gan2016,Fu2016,Wei2017,Gao2018,Si2019,Cohen-Stead2019}.
The application of strain could alter the band structure and thus provides the specific insight into the nature of a CDW phase, such as exploring the nature of Fermi surface nesting~\cite{Fast1998,Johannes2008}.
Unusual phenomena may emerge at the quantum critical points of strain-induced structural quantum phase transitions~\cite{Gruner2017}.
From the experimental point of view, strain is one of the few available handles that could be utilized to controllably and reversibly tune electronic structures.
In bulk single crystal, strain can be applied by attaching materials to piezoelectric substrates.
While in thin films, strain can be generated by using the lattice mismatch between the film and the substrate.

For uranium, the epitaxial thin film, in which the strain is present, has already led to the discovery of a wide range of new structural, magnetic, and electronic phenomena~\cite{Molodtsov1998,Berbil-Bautista2004,Ward2008,Springell2008,Springell2014,Chen2019}.
The hcp phase of uranium, which does no exist in the bulk, could be stabilized on Gd/Nb substrate~\cite{Springell2008}.
All the allotropic phases of uranium are known as non-magnetic and confirmed by the theoretical calculation, while hcp-U is predicted to order magnetically.
In addition, the CDW transition temperature increase from 43 K in the bulk to 120 K in the thin film of uranium on W~\cite{Springell2014}.
Similar behavior could be found in the other CDW materials~\cite{Xi2015,Chen2015,Chen2016}.
From the angle-resolved photoemission spectroscopy of uranium films on W(110), large spectral weight was observed around the Fermi level~\cite{Chen2019}.

\begin{figure}[t!]
\centering
\includegraphics[width=0.59\columnwidth]{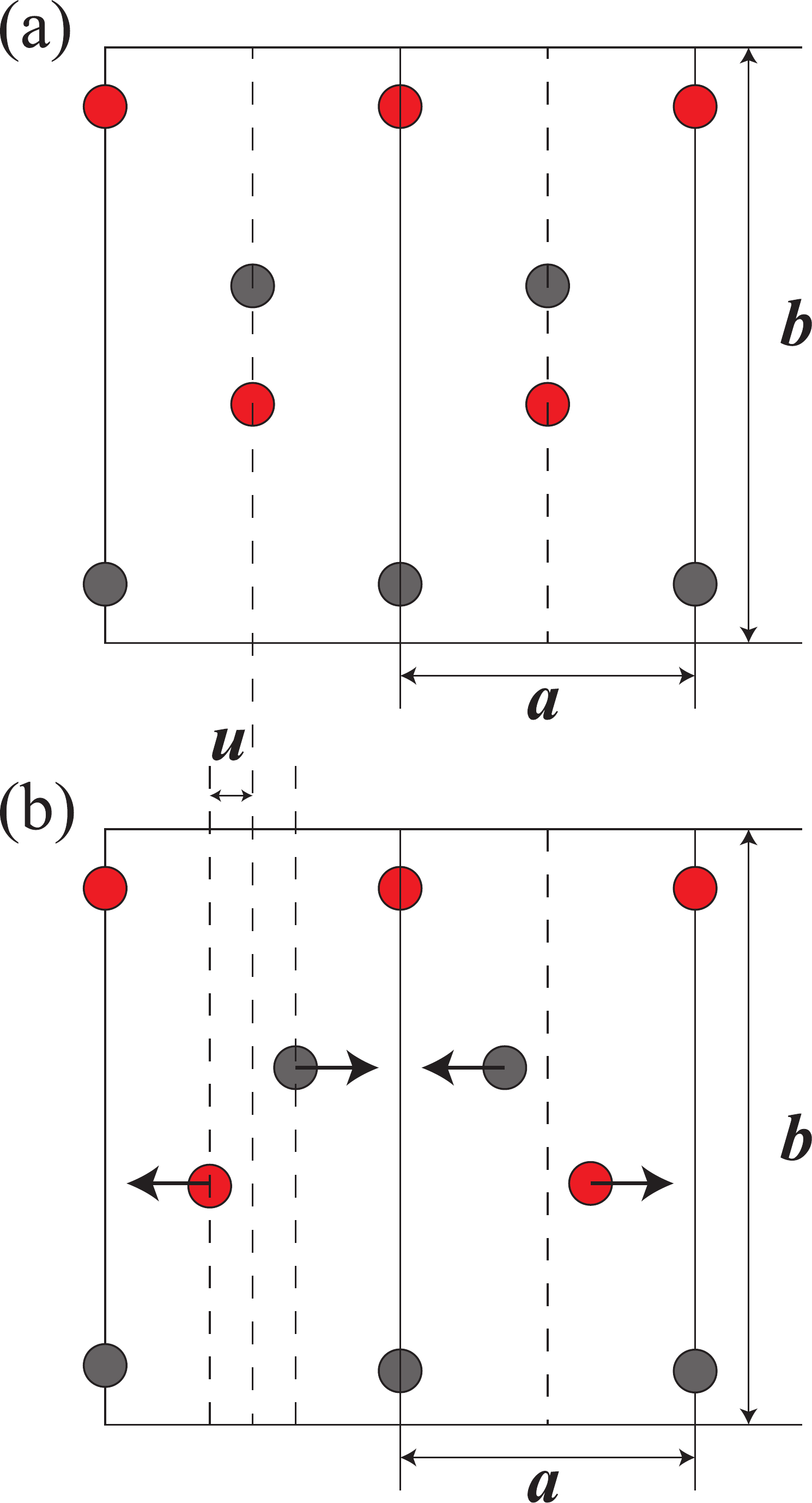}
\caption{The crystal structures of $\alpha$-U (a) and $\alpha_{1}$-U (b) used in the calculations. Both are shown in the $ab$ plane. Note that the CDW distortion is determined by the displacement $u$. The gray (red) circles mark atoms situated in the $z$ = $\textstyle{\frac{1}{4}}$ ($z$ = $\textstyle{\frac{3}{4}}$) layer.}
\label{fig:structure}
\end{figure}

Most of the theoretical work on the CDW of uranium has been within first-principles density-functional theory (DFT)~\cite{Fast1998,Bouchet2008,Raymond2011,Springell2014,Qiu2016} and focus on the transition from $\alpha$ to $\alpha_1$.
The pioneering work~\cite{Fast1998} constructed the model of $\alpha_1$-U from displacing the atoms in the $2\times1\times1$ supercell of $\alpha$-U (See Fig.~\ref{fig:structure}) and employed the full-potential linearized augmented plane wave (FP-LAPW) method to evaluate the energy difference between $\alpha$-U and $\alpha_1$-U.
It was found that the small displacement ($\sim$0.028\AA) is energetically favorable and the energy gain is in agreement with the CDW transition temperature 43 K.
In particular, the structural distortion could be interpreted using the nesting features of the Fermi surface in $\alpha$-U.
Phonon calculations using pseudopotential method also show that the energy of the $\Sigma_4$ phonon mode along [100] direction varies between negative and position for different pressure~\cite{Raymond2011} and strain~\cite{Springell2014}.

The main purpose of this work is a systematic investigation of the effect of the uniaxial strain on the CDW order in $\alpha$-U from first-principles calculation in an attempt to establish a primary connection between strain and CDW order.
It is found that the CDW charge order could be significantly enhanced by the tensile strain along $a$ or $b$ direction, and particularly the compressive strain along $c$ direction.
The rest of this paper is organized as follow.
In Sec.~\ref{sec:computational}, we present the computational details.
Section~\ref{sec:result} describes the results and gives our discussion. 
Our main conclusion is summarized in Sec.~\ref{sec:conclusion}. 

\section{Computational details}
\label{sec:computational}

The calculation of electronic structure, total energy, and force is performed within the framework of DFT~\cite{Hohenberg1964,Kohn1965} using the Perdew–Burke–Ernzerhof generalized gradient approximation~\cite{Perdew1996}.
The inclusion of relativistic effect is treated using spin-orbit coupling.
The atomic structures in Fig.~\ref{fig:structure} are constructed from the experimental lattice parameters~\cite{Smith1980,Fast1998} and used in the calculations.
Here the qualitative conclusion on the strain effect on CDW order is independent of the lattice parameters.
Note that a $2\times1\times1$ supercell of conventional cell is used here for $\alpha$-U to keep the number of atoms being equal to that in $\alpha_1$-U.
To solve the Kohn-Sham equation, both the FP-LAPW method~\cite{Singh2006} and the projected augment wave (PAW) pseudopotential scheme~\cite{Bloechl1994,Kresse1999} are used.
On the one hand, the FP-LAPW method is applied here to accurately calculate the total energy and electronic structure.
On the other hand, the evaluation of the force and the related phonon mode, which is the shortcoming of FP-LAPW method, is carried out using the PAW pseudopotential scheme.

The FP-LAPW method is implemented in the \verb|WIEN2k| code~\cite{Blaha1990}.
The cutoff parameter is $R_{\rm MT}K_{\rm MAX}$ = 11.0, and the muffin-tin radius for the U atom is fixed at 2.45 bohr.
The Brillouin zone integration is done on a $13\times16\times13$ mesh, resulting in 343 $k$ points in the irreducible wedge of the first Brillouin zone.

The \verb|VASP| code~\cite{Kresse1996} is used to employ the pseudopotential calculation, in which the plane wave basis is taken as 600 eV and the Brillouin zone integration is based on $11\times11\times11$ $k$ point meshes. 
The total energies and Fermi surfaces from PAW calculations are compared with those from FP-LAPW (see Appendix~\ref{sec:appendix}).
The force constants are calculated within the framework of density-functional perturbation theory~\cite{Gonze1997} and the phonon modes are obtained from the postprocessing using the \verb|phonopy| package~\cite{Togo2015}.

\begin{figure}[t!]
\centering
\includegraphics[width=0.99\columnwidth]{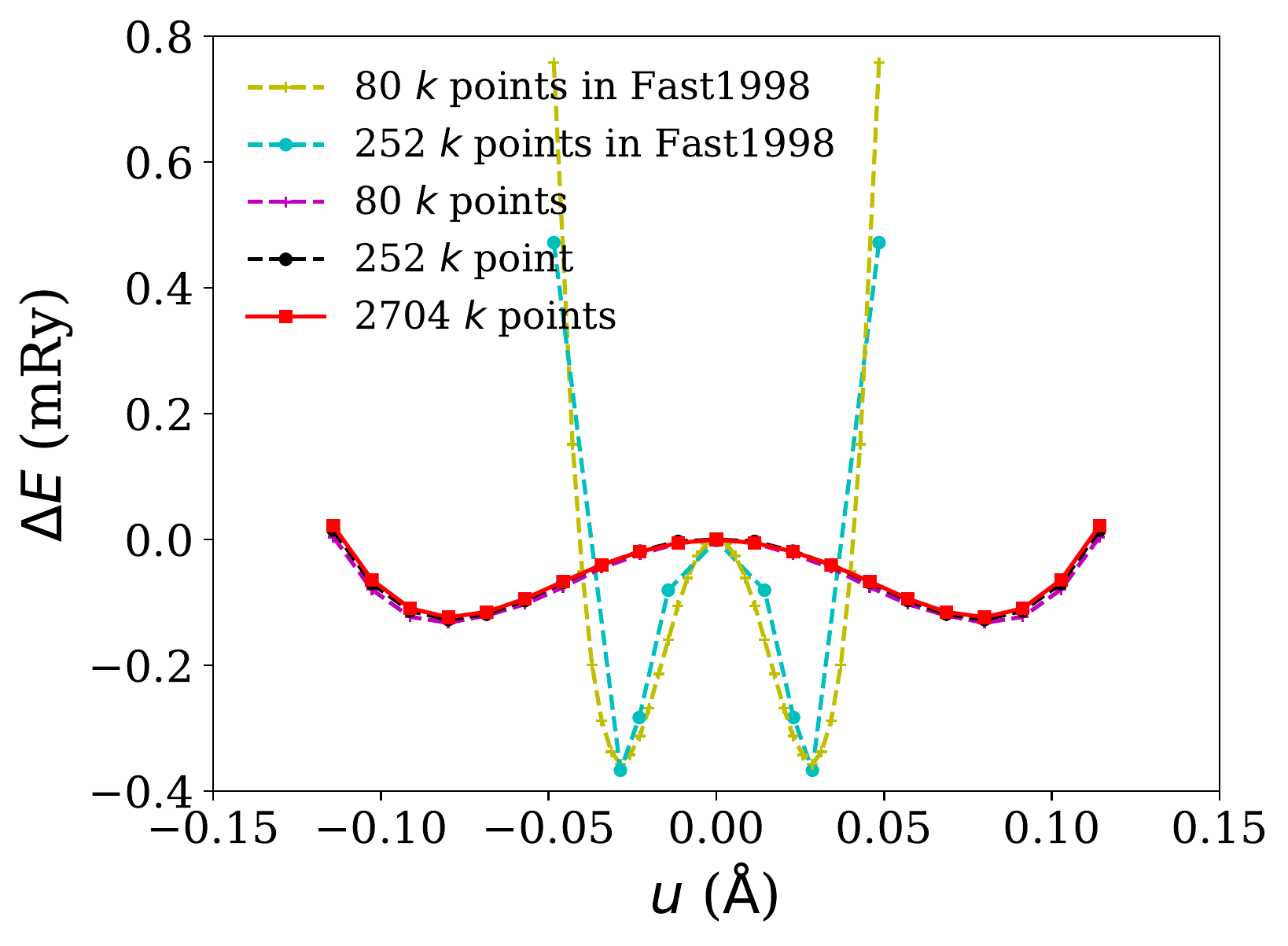}
\caption{The calculated total energy difference per atom $\Delta E$ (\ref{eq:energy}) as a function of the displacment parameter $u$ in Fig.~\ref{fig:structure}. Three different $k$ points were used and the results from Ref.~\cite{Fast1998} are replotted here.}
\label{fig:energy}
\end{figure}

\section{Results and discussion}
\label{sec:result}

From the perspective of experiment, the most straightforward and measurable effect of strain on the CDW order in $\alpha$-U is the modification of transition temperature $T_{\rm CDW}$. 
Since the phase transition from $\alpha$ to $\alpha_1$ is recognized to be first order~\cite{Steinitz1970}, $T_{\rm CDW}$ is closely related to energy difference between $\alpha_1$-U and $\alpha$-U~\cite{Fast1998}, i.e., the CDW formation energy, 
\begin{eqnarray}
\Delta E = E_{\alpha-{\rm U}} - E_{\alpha_1-{\rm U}},\label{eq:energy}
\end{eqnarray}
in which $E_{\alpha-{\rm U}}$ and $E_{\alpha_1-{\rm U}}$ are the total energies per atom of $\alpha$-U and $\alpha_1$-U, respectively.
Figure~\ref{fig:energy} plots the energy difference $\Delta E$ as a function of distortion $u$. 
The double-well shape, which is found in Ref.~\cite{Fast1998} and reproduced in our calculation, indicates that $\alpha_1$-U with finite distortion is stable with respect to undistorted $\alpha$-U.
Our calculations are consistent with the results from Ref.~\cite{Fast1998} not only in the energy difference but also in the Fermi surface, which is shown in Fig.~\ref{fig:fermi} (see Appendix~\ref{sec:appendix}).
The topological change of the Fermi surface close to the point $U$ ($\textstyle{\frac{1}{2}}$,0,$\textstyle{\frac{1}{2}}$) in the reciprocal space is also shown here.
In addition, let us note that the magnitude of $u$ to minimize the total energy $u_{\rm max}$ $\sim$ 0.08~\AA~and the corresponding energy change $\Delta E_{\rm min}$ $\sim$ -0.13 mRy per atom. 
This quantitative discrepancy from that from Ref.~\cite{Fast1998} may result from the different computational parameters or lattice parameters.

\begin{figure}[t!]
\centering
\includegraphics[width=0.95\columnwidth]{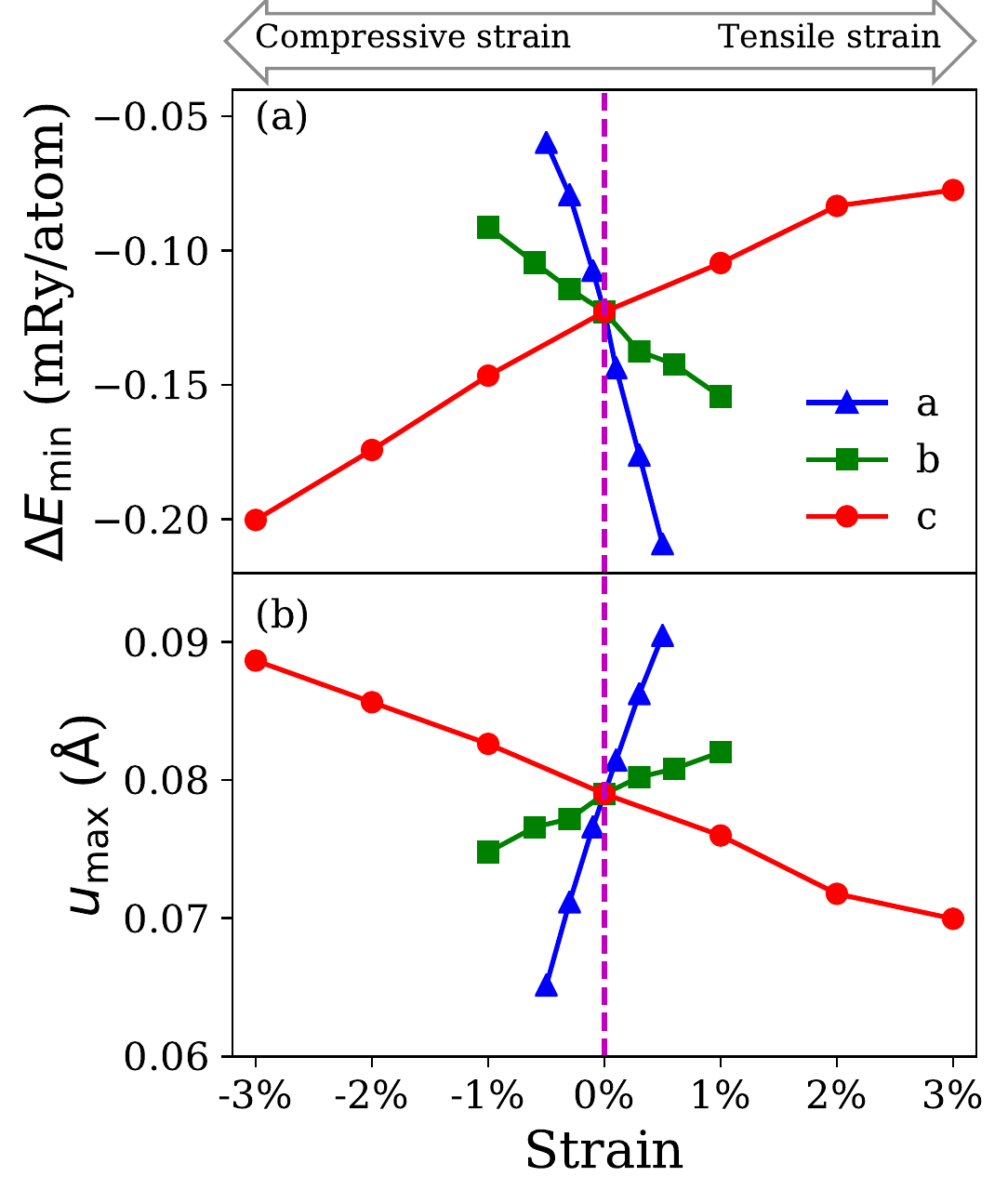}
\caption{(a) The minimum CDW formation energy $\Delta E_{\rm min}$ and (b) maximum displacement $u_{\rm max}$ with respect to the uniaxial strain along $a$, $b$, and $c$ directions.}
\label{fig:strain}
\end{figure}

Now let us investigate the evolution of the CDW order of $\alpha$-U under the uniaxial compressive and tensile strains.
Figure~\ref{fig:strain} shows the minimum CDW formation energy and the corresponding atomic displacement in the CDW phase as a function of uniaxial strain along $a$, $b$, and $c$ directions.
When the tensile strain is applied along $a$ and $b$ directions, the formation energy $\Delta E_{\rm min}$ gradually decreases and the maxmium displacement $u_{\rm max}$ increases, indicating that the CDW phase becomes more stable and the CDW transition temperature $T_{\rm CDW}$ increases.
In contrast, the compressive strain along $a$ and $b$ directions gradually suppresses the CDW stability and $T_{\rm CDW}$ is very likely to decrease.
This kind of strain effect, which is similar to the effect of pressure~\cite{Lander1994,Soderlind1995,Fast1998,Qiu2016}, can be easily understood. 
By analogy, the effect of tensile strain is equivalent to that of negative pressure and the transition temperature would be lift.
The compressive strain is analogous to the positive pressure and suppress the CDW transition.
From the viewpoint of one-dimensional atomic chain~\cite{Johannes2008}, the tensile strain along $a$ direction would increase the atomic distance and favors the Peierls transition.

\begin{figure}[t!]
\centering
\includegraphics[width=0.95\columnwidth]{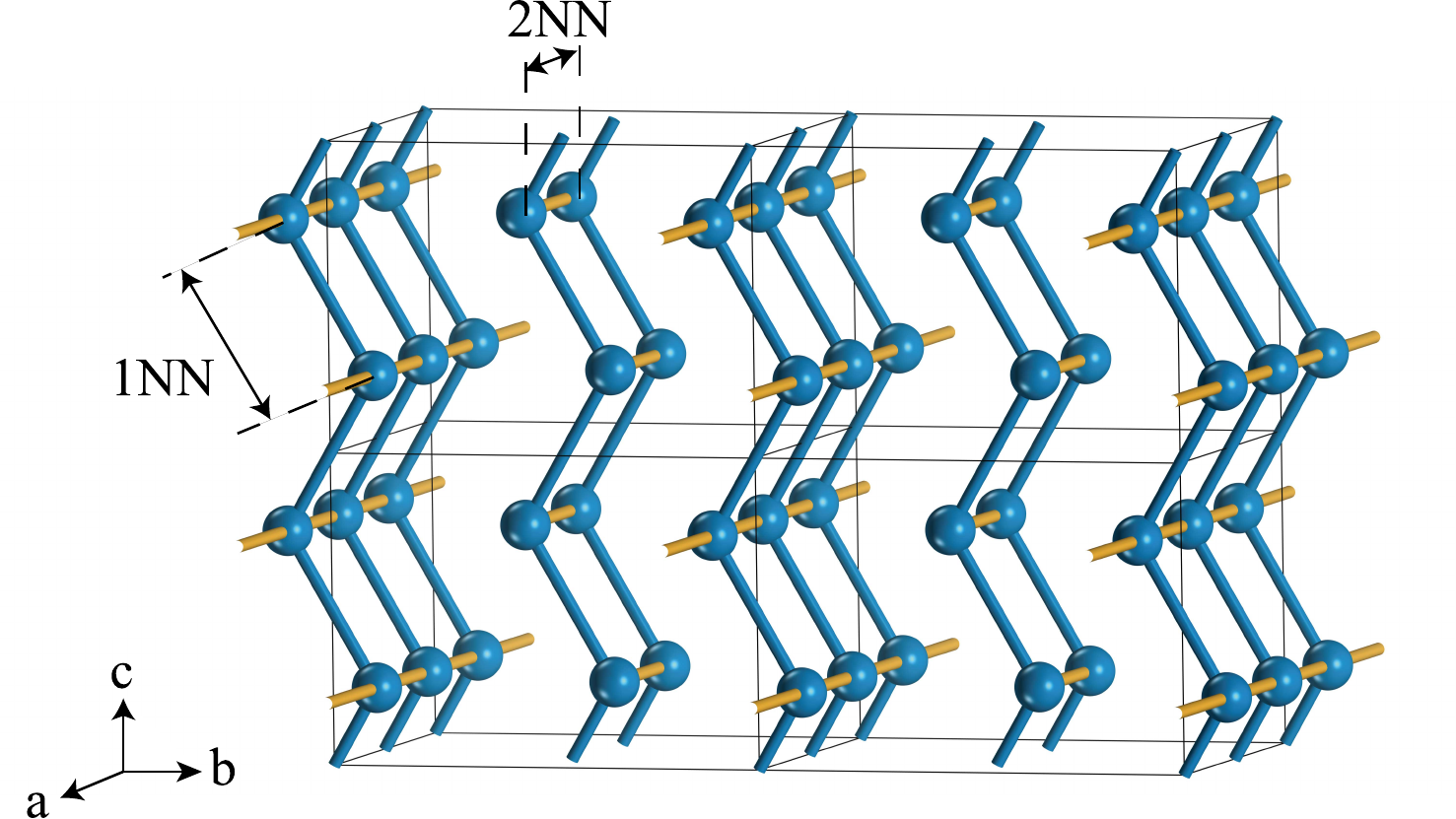}
\caption{The $\alpha$-U structure in a 2$\times$2$\times$2 supercell in which the first nearest-neighbor (1NN, blue) and second nearest-neighbor (2NN, gold) bonds are shown.}
\label{fig:zigzag}
\end{figure}

Opposite behavior occurs for the strain along $c$ direction.
In this case, the CDW formation energy $\Delta E$ increases with the increase of the tensile strain and decreases when the compressive strain is applied.
This means that the CDW stability is suppressed by the tensile strain along $c$ direction rather than the compressive strain along $c$ strain.
This result is quite counter-intuitive and not easily understood.
Here let us first attempt to discuss this anomalous behavior from the perspective of atomic structure.
In Fig.~\ref{fig:zigzag}, the $\alpha$-U structure in 2$\times$2$\times$2 supercell is presented.
The structure can be conceived as a group of corrugated rectangular layers stacked in the $b$ direction.
The degree of corrugation is controls by the angle $\theta_{\rm c} = {\rm arctan}(4yb/c)$, which is a monotonically decreasing function of $c$ for positive $4yb/c$. 
As the tensile strain along $c$ direction is increased, the degree of corrugation decrease, resulting the formation of one-dimensional atomic chain along $c$ direction.
This atomic chain would affect the pre-exist atomic chain along $a$ direction (see Fig.~\ref{fig:structure}) where the CDW transition into $\alpha_1$-U occurs. 
The spacing of the new and pre-exist atomic chain is first and second nearest-neighbor distance, respectively.
This shorter interatomic distance means that the formation of atomic chain along $c$ direction yields the decrease of the tendency of CDW transition along $a$ direction. 
Therefore, the tensile strain along $c$ direction make the CDW unstable and the compressive strain along $c$ direction stabilize the CDW phases.

Another perspective of atomic structure could be obtained from the variation of experimental lattice constants in $\alpha$-U at low temperature~\cite{Barrett1963,Steinitz1970}. 
As the temperature is decreased, the lattice constants $a$ and $b$ first decrease to a minimum at 43 K and then rapidly increase while $c$ decreases more rapidly below this temperature. 
That is to say, the cell dimensions $a$ and $b$ should expand and $c$ should contract for the stabilization of CDW order.
This experimental observation is in agreement with our theoretical evaluation of strain effect on the CDW order.

Figure~\ref{fig:strain} also indicates that one can use the strain to effectively tune the CDW order.
A very small strain, such as 0.5\% tensile strain along $a$ direction, would double the CDW formation energy and is very likely to increase the transition temperature $T_{\rm CDW}$.
This magnitude of strain is easily accessible in the experiment through the lattice mismatch between the substrate and the uranium film~\cite{Springell2014}.
Compared to the layered transition-metal chalcogenide such as 1$T$-TiSe$_2$, the magnitude of strain to take effect is much smaller, which illustrate the sensitivity of 5$f$ electrons.

\begin{figure}[t!]
\centering
\includegraphics[width=0.95\columnwidth]{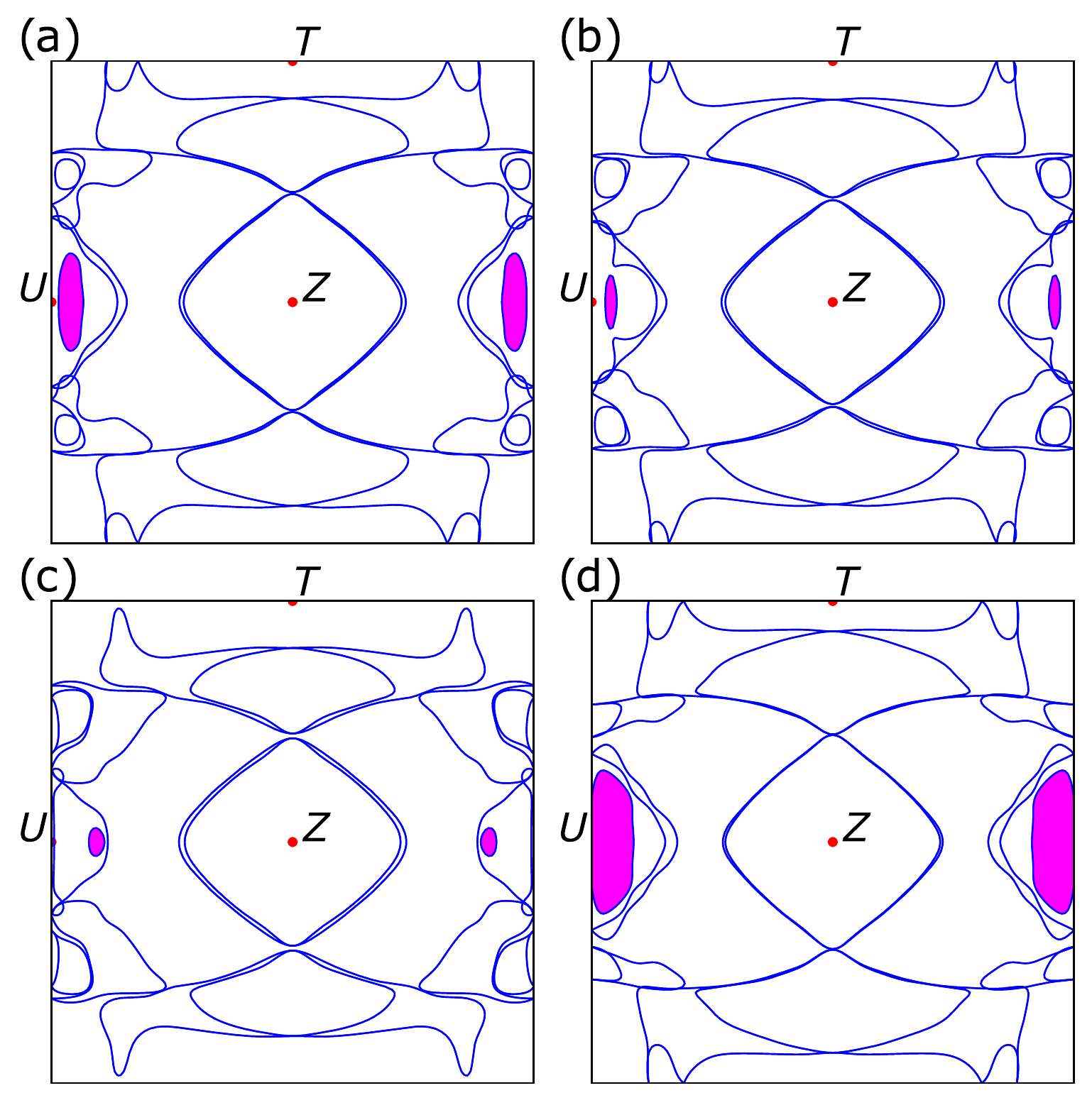}
\caption{The Fermi surface of $\alpha$-U when (a) -0.5\% and (b) 0.5\% strains along $a$ direction, (c) -3\% and (d) 3\% strains along $c$ direction are applied. The ellipsoid objects are highlighted in violet. The symmetry points include $Z$ (0, 0, 1/2), $T$ (0, 1/2, 1/2) and $U$ (1/2, 0, 1/2).}
\label{fig:surface}
\end{figure}

The effect of strain on the CDW order could also be viewed from the perspective of electronic structure.
Let us go back to the topological change of Fermi surface around $U$ point in Fig.~\ref{fig:fermi}.
When comparing the Fermi surfaces of $\alpha$-U and $\alpha_1$-U, one can find that an ellipsoid object disappears due to the CDW distortion.
It is equivalent to the band splitting at the Fermi level at the corresponding $k$ points, i.e., Peierls distortion. 
The smaller the ellipsoid object is, the easier the Peierls distortion occurs.
In Fig.~\ref{fig:surface}, we plot the Fermi surfaces of $\alpha$-U when the tensile and compressive strains along $a$ and $c$ directions are applied.
Clearly, the ellipsoid object is very small for the tensile strain along $a$ direction and compressive strain along $c$ direction.
On the contrary, this object become larger for the compressive strain along $a$ direction and tensile strain along $c$ direction.
These electronic structure calculations reconfirm that the effect of strain along $c$ direction on the CDW order in $\alpha$-U is abnormal.

\begin{figure}[t!]
\centering
\includegraphics[width=0.95\columnwidth]{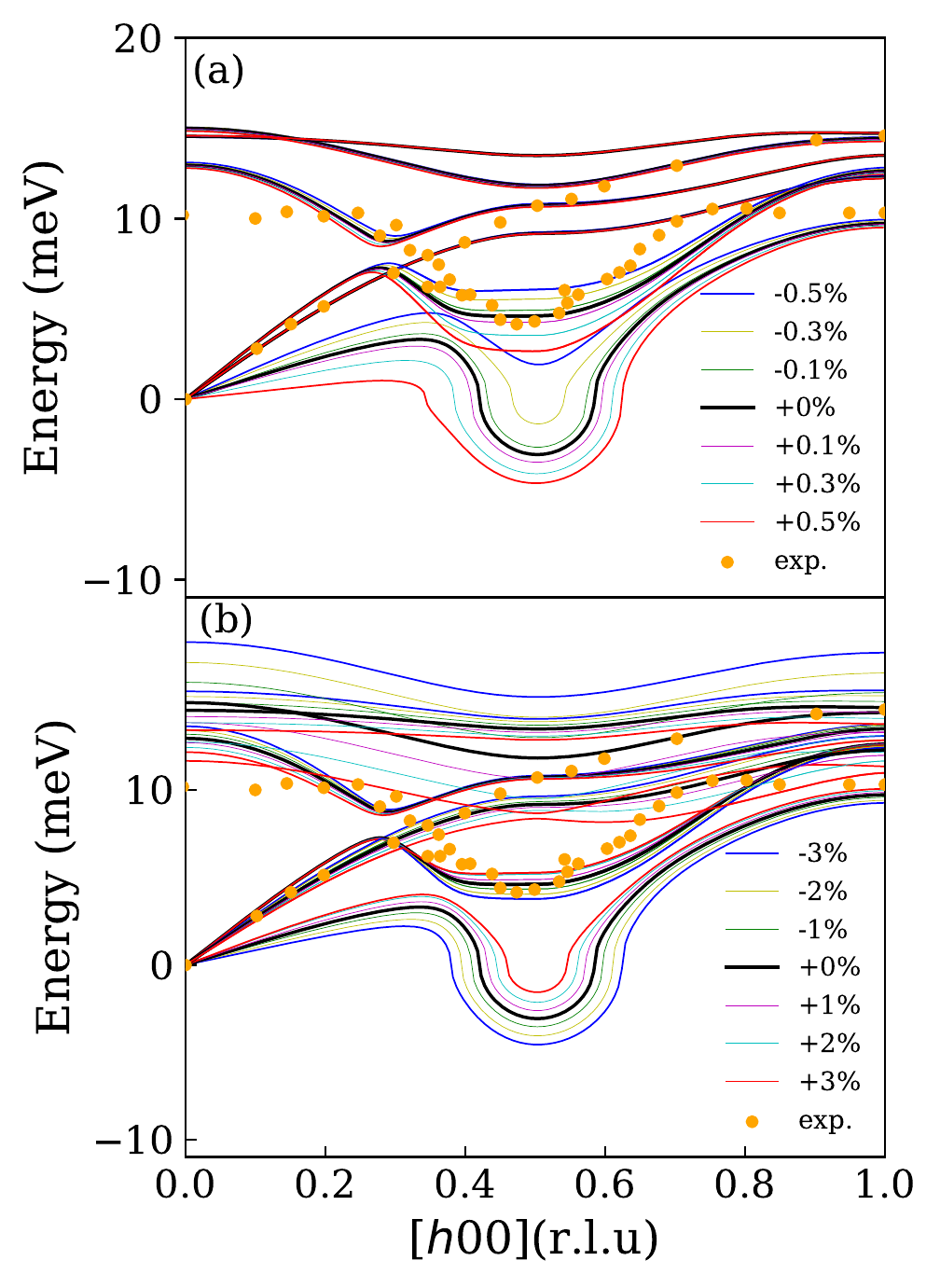}
\caption{The phonon modes of $\alpha$-U when strain along (a) $a$ and (b) $c$ directions are applied. THe experimental points at room temperature for key modes are shown~\cite{Lander1994}.}
\label{fig:phonon}
\end{figure}

Furthermore, the energetical instability usually indicates an intrinsic soft mode in $\alpha$-U, i.e., the dynamical instability that drives the formation of the CDW transition~\cite{Raymond2011}.
The $\alpha$ phase of uranium (see Figs.~\ref{fig:structure} and \ref{fig:zigzag}) is unstable at zero temperature, as demonstrated by the results in Fig.~\ref{fig:phonon} (black solid line).
Our calculation is consistent with the experimental data at room temperature~\cite{Lander1994}.
From the other calculations in Fig.~\ref{fig:phonon}, $\alpha$-U is even more stable for the compressive strain along $a$ directions and tensile strain along $c$ direction. 
In particular, $\alpha$-U is dynamically stable for a compressive strain of -0.5\%, which is in agreement with the previous theoretical computation~\cite{Springell2014}.
In contrast, $\alpha$-U is even more unstable for the tensile strain along $a$ direction and compressive strain along $c$ direction, 
Overall, our argument about the abnormal effect of strain along $c$ direction is also supported by the phonon calculations.

\section{Conclusion}
\label{sec:conclusion}

In summary, we investigate the effect of the uniaxial strain on the charge density wave (CDW) order in $\alpha$-U within the framework of relativistic density-functional theory.
When the strains along $a$, $b$, and $c$ are applied, the total energy, Fermi surface, and phonon mode are compared between the $\alpha$-U structure with and without CDW distortion. 
From the calculation of the total energy difference, it is found that the tensile strain along $a$/$b$ direction increase the CDW instability and make $\alpha$-U even more unstable.
The compressive strain along $a$/$b$ direction takes the opposite effect.
This is very similar to the effect of pressure, in agreement with the previous literature, and intuitive.
On the contrary, the effect of strain along $c$ direction is counter-intuitive, in which the compressive/tensile strain increases/decreases the energetical instability of $\alpha$-U.
This could be understood from the perspective of atomic, electronic, and phonon structures.
From the atomic structure of $\alpha$-U, the increase of cell dimension $c$ means the emergence of a new one-dimensional atomic chain which is the rival of previous one-dimensional atomic chain for CDW distortion.
The calculated Fermi surfaces and phonon modes for different strain also present the abnormal effect of strain along $c$ direction with respect to that of strain along $a$/$b$ direction.

It is noteworthy that only the first CDW transition in $\alpha$-U is considered here.
Apparently, the strain would take effects on the other two CDW transitions. 
But the supercell for the structures with more CDW transitions is too huge and unreachable for modern supercomputer.
In addition, the biaxial, triaxial, and more complex strains should take place in the experiment and the researches would be pursued in the near future.

\section*{Acknowledgements}

The work was supported by the National Natural Science Foundation of China (Grant Nos. 22176181, 11874306, and 12174320), the Foundation of Science and Technology on Surface Physics and Chemistry Laboratory (Grant No. WDZC202101), and the Natural Science Foundation of Chongqing (Grant No. cstc2021jcyj-msxmX0209).

\bibliography{cdw-strain}

\appendix
\section{Appendix}
\label{sec:appendix}

\begin{figure}[t!]
\centering
\includegraphics[width=0.8\columnwidth]{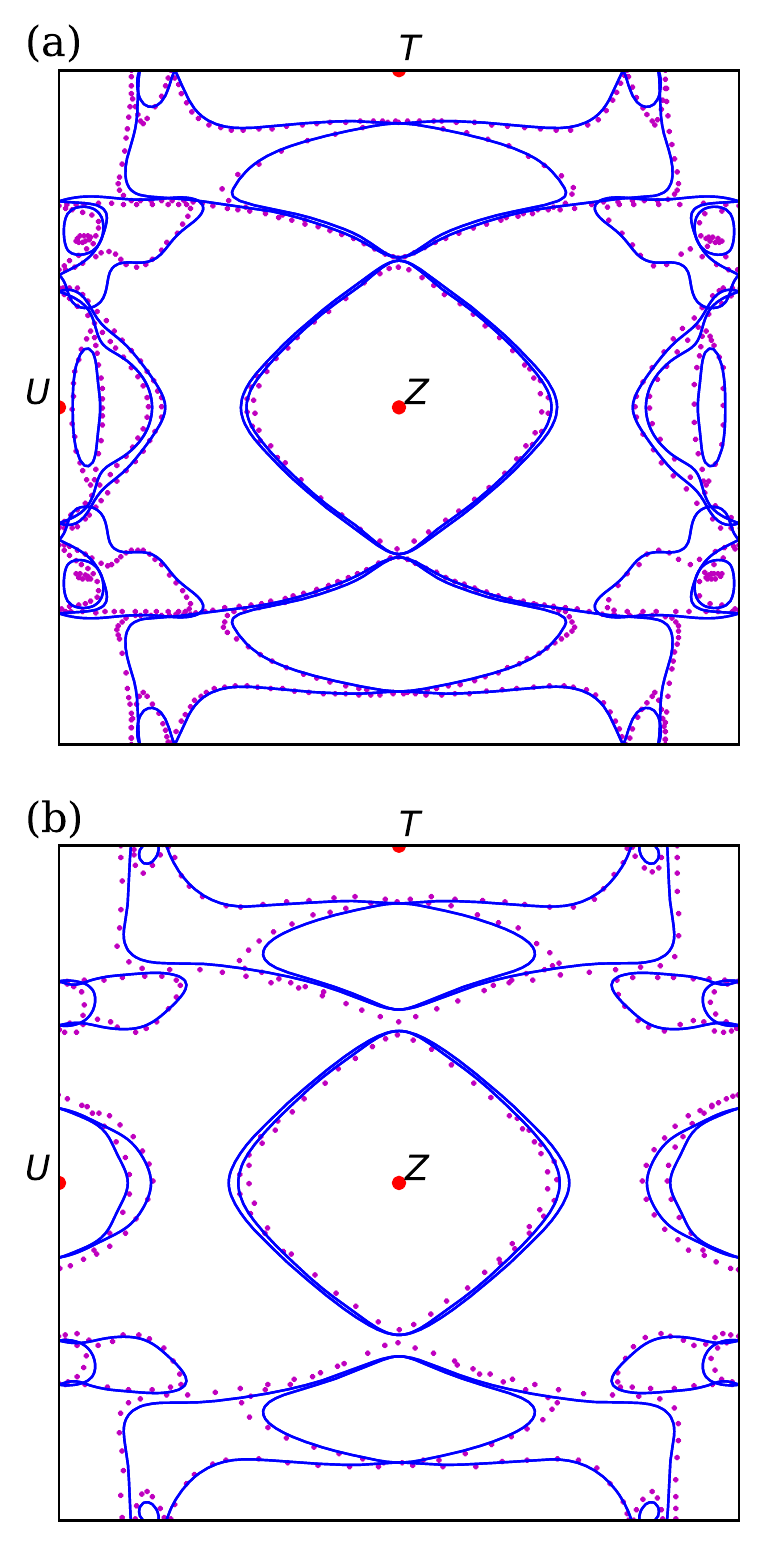}
\caption{The Fermi surface of (a) $\alpha$-U and (b) $\alpha_{1}$-U with $u$ = 0.08~\AA~in the $k_z$ = 1/2 plane. 
The dotted line are replotted using the figure in Ref.~\cite{Fast1998}.
The symmetry points includes $Z$ (0, 0, 1/2), $T$ (0, 1/2, 1/2) and $U$ (1/2, 0, 1/2).}
\label{fig:fermi}
\end{figure}

In this appendix, the supporting materials are provided.
In Fig.~\ref{fig:fermi}, the Fermi surface of $\alpha$-U with and without distortion is presented.
Our calculation is compared with that from the literature and the agreement is good.
In particular, the topological change around the $U$ point is very clear.

\begin{figure}[t!]
\centering
\includegraphics[width=0.9\columnwidth]{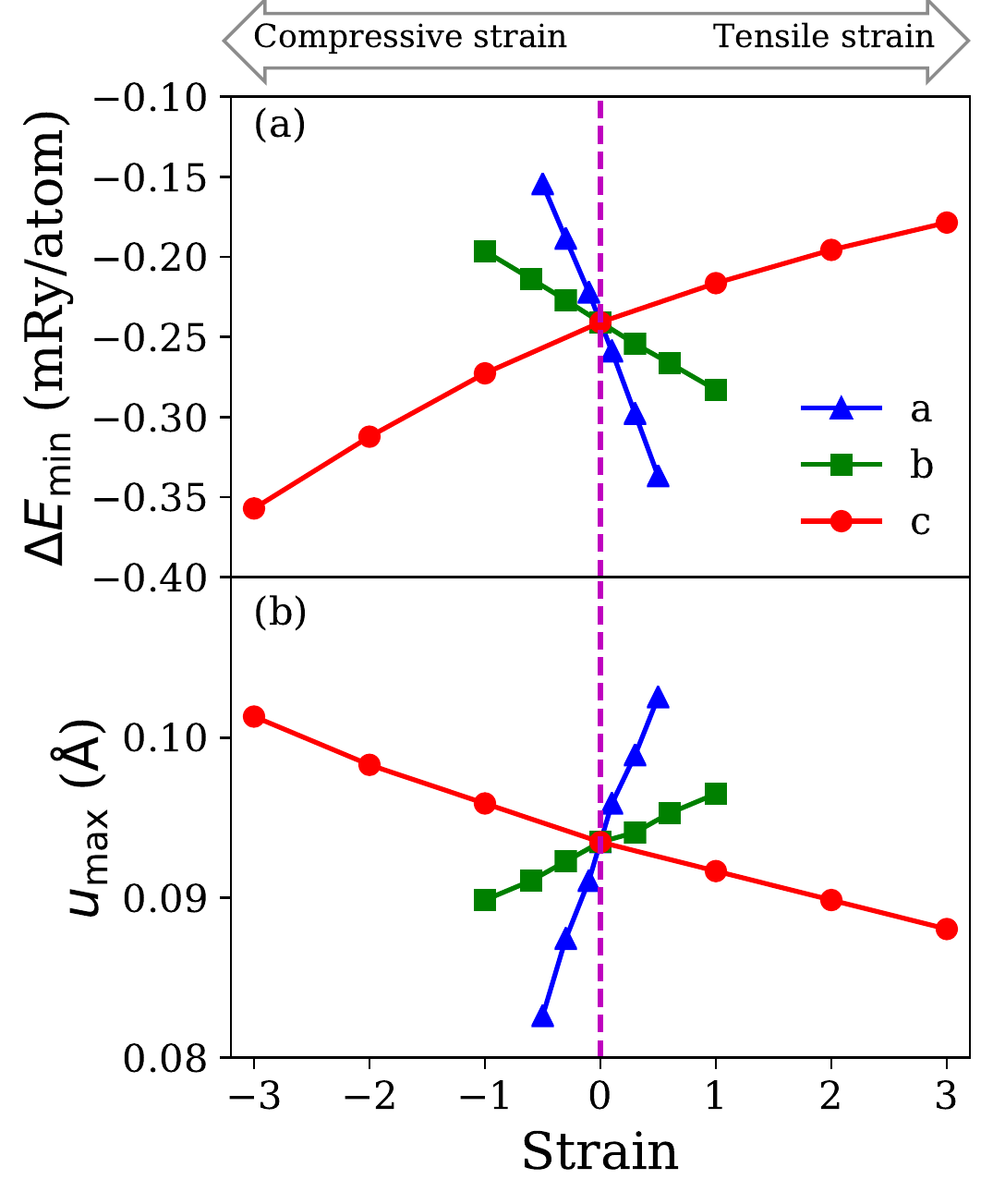}
\caption{(a) The CDW formation energy $\Delta E_{\rm min}$ and (b) maximum displacement $u_{\rm max}$ with respect to the uniaxial strain along $a$, $b$, and $c$ directions from PAW calculations.}
\label{fig:strain-paw}
\end{figure}

\begin{figure}[t!]
\centering
\includegraphics[width=0.9\columnwidth]{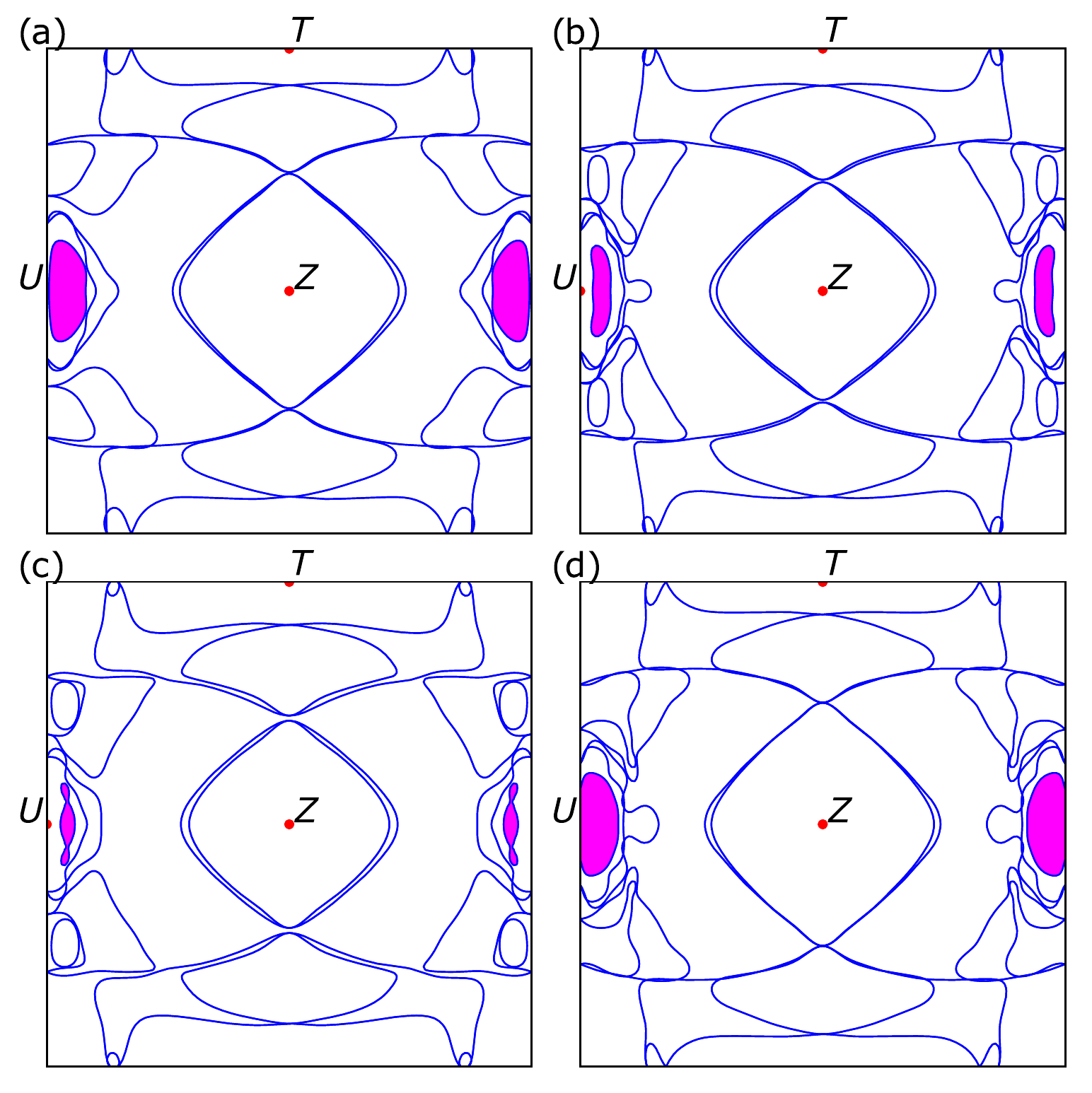}
\caption{The Fermi surface of $\alpha$-U from PAW calculations. Here (a) -0.5\% and (b) 0.5\% strains along $a$ direction, (c) -3\% and (d) 3\% strains along $c$ direction are applied. The ellipsoid objects are highlighted in violet. The symmetry points includes $Z$ (0, 0, 1/2), $T$ (0, 1/2, 1/2) and $U$ (1/2, 0, 1/2).}
\label{fig:surface-paw}
\end{figure}

For a evaluation of PAW scheme, the CDW formation energy and distortion magnitude are also calculated using \verb|VASP| and shown in Fig.~\ref{fig:strain-paw}. The Fermi surface of $\alpha$-U with different strains is presented in Fig.~\ref{fig:surface-paw}. The results from PAW calculations are in agreement with those from LAPW calculations.

\end{document}